\documentclass[referee]{aa} % for a referee version
\usepackage{graphicx}
\usepackage{txfonts}
\begin{document}
\def\xr#1{\parindent=0.0cm\han indent=1cm\han after=1\indent#1\par}
\def\la{\raise.5ex\hbox{$<$}\kern-.8em\lower 1mm\hbox{$\sim$}}
\def\ma{\raise.5ex\hbox{$>$}\kern-.8em\lower 1mm\hbox{$\sim$}}
\def\ea{\it et al. \rm}
\def\am{$^{\prime}$\ }
\def\as{$^{\prime\prime}$\ }
\def\msol{M$_{\odot}$ }
\def\kms{$\rm km\, s^{-1}$}
\def\cm3{$\rm cm^{-3}$}
\def\Ts{$T_{*}$}
\def\Vs{$V_{s}$}
\def\n0{$n_{0}$}
\def\B0{$B_{0}$}
\def\ne{$n_{e}$}
\def\Te{$T_{e}$}
\def\T r{$T_{ r}$}
\def\T as{$T_{ as}$}
\def\Ec{$E_{c}$}
\def\Fh{$F_{H}$}
\def\erg{$\rm erg\, cm^{-2}\, s^{-1}$}
\def\Hb{H$\beta$}
\def\Ha{H$\alpha$}

   \title{The Narrow Line Region of Ark\,564}

   \subtitle{}

   \author{Marcella Contini
          \inst{1,3}
          \and
          Alberto Rodr\'{\i}guez-Ardila
          \inst{2}
          \and
          Sueli Viegas
          \inst{1}
          }

   \offprints{M. Contini}

   \institute{Instituto de Astronomia, Geof\'{\i}sica e Ci\^encias
Atmosf\'ericas. Rua do Mat\~ao 1226. CEP 05508-900, S\~ao Paulo, SP,
Brazil.\\
              \email{marcel@wise1.tau.ac.il, viegas@astro.iag.usp.br}
         \and
             Laborat\'orio Nacional de Astrof\'{\i}sica $-$ Rua dos
Estados Unidos 154 $-$ Bairro das Na\c c\~oes. CEP 37500-000, Itajub\'a,
MG, Brazil. \\
             \email{aardila@lna.br}
         \and
             School of Physics and Astronomy, Tel-Aviv University,
Ramat-Avit, Tel-Aviv, 69978, Israel.      
             }

   \date{Received; accepted}

   \abstract{
The continuum and emission-line spectrum of the 
narrow-line Seyfert 1 galaxy  Ark\,564 is used to 
investigate, for the first time, the physical 
conditions and structure of its narrow line 
region (NLR). For this purpose, composite 
models, accounting for the coupled effect of 
photoionization and shocks, are employed. The 
emission-line spectrum of Ark\,564, which ranges 
from the ultraviolet to the near-infrared, shows
a rich forbidden line spectrum. Strong emphasis
is given to the study of the coronal line region.
The diversity of physical conditions deduced 
from the observations requires multi-cloud 
models to reproduce the observed 
lines and continuum. We find that 
a combination of high velocity (\Vs = 1500  \kms) 
shock-dominated clouds as well as
low velocity ( \Vs = 150 \kms) radiation-dominated clouds 
explains the coronal lines, while the optical low-ionization lines 
are mainly explained by shock-dominated clouds.
The results for Ark\,564 are compared with those obtained
for other Seyfert galaxies previously analyzed such as NGC\,5252, Circinus,
NGC\,4051 and NGC\,4151.
 The model 
results for the ultraviolet and optical permitted lines suggest
that the broad line region may contribute up to 80\%, 
depending on the emission-line, being of about 30\% for H$\beta$. 
The consistency of the multi-cloud model is
checked by comparing the predicted and observed continuum, 
from radio to X-ray, and indicate
that the dust-to-gas ratio in the clouds varies from 10$^{-15}$ to 10$^{-12}$.
   }

   \maketitle
%
%________________________________________________________________

\section{Introduction}

Narrow-line Seyfert 1 galaxies (NLS1) are a peculiar
group of active galactic nuclei (AGN) with extreme
properties. Their permitted optical lines emitted by 
the broad line region (BLR) show full widths at 
half-maximum (FWHM) not exceeding 2000 \kms, 
the [O\,{\sc iii}]/H$\beta$ ratio is less than 3, and 
the UV-optical spectrum is usually dominated by strong 
Fe\,{\sc ii} emission multiplets (Osterbrock \& Pogge 
\cite{op85}). In the soft X-ray, NLS1 have generally 
much steeper continuum slopes and rapid variability 
(Boller et al. \cite{bbf96}). 
In addition, Leighly (\cite{l99}) found that the 
hard X-ray photon index is significantly steeper in NLS1 
compared with that of classical Seyfert 1 and that 
soft excess emission appears considerably more frequently 
in the former than in the later. Nowadays, the most
accepted explanation for the above properties
is that these objects have relatively low black-hole 
masses for their luminosity but high accretion rates 
(Boller et al. \cite{bbf96}). In fact, observational 
evidence obtained through reverberation
mapping techniques points toward the presence of a 
black hole of smaller mass in NLS1 (M$_{\rm BH} \sim$ 
10$^{6-7}$ \msol) than that of  other broad-line AGN 
(M$_{\rm BH} \sim$ 10$^{7-9}$ \msol;
Peterson \& Wandel \cite{pw00}) with similar luminosity.

While most efforts have been devoted to explain the
extreme properties associated to the BLR in NLS1,
very little has been said about the properties of
the narrow line region (NLR) in these objects and
how do they compare to that of other AGN, except for a
few works reported in the literature (Goodrich \cite{good89};
Rodr\'{\i}guez-Ardila et al. \cite{ro00}; Nagao et al. \cite{na01};
Rodr\'{\i}guez-Ardila et al. \cite{ro02}). Given that the NLR is one of the 
fundamental ingredients of AGN, every effort to
understand what NLS1 are must include a detailed
investigation of the NLR at least for 
the following three reasons. 

First, the NLR is the
largest structure directly affected by the radiation
from the central engine. Therefore, if the spectral energy 
distribution (SED) of the ionizing radiation flux is somehow
different from that of classical Seyferts, the resulting 
ionization structure of the NLR gas should be different, 
as it is observed in some NLS1 (Rodr\'{\i}guez-Ardila et al.
\cite{ro00}). 

Second, NLS1 are located at the 
extreme end of the anticorrelation between [O\,{\sc iii}] 
and Fe\,{\sc ii} (Boroson \& Green \cite{bg92}), namely, 
Fe\,{\sc ii} is strong in objects with weak 
[O\,{\sc iii}] emission and weak in those with strong
[O\,{\sc iii}]. That inverse correlation suggests the
existence of a physical mechanism which balances Fe\,{\sc ii} 
excitation against the illumination of the NLR. Although 
the nature of the Fe\,{\sc ii} is out of the scope of this 
paper, the study of the NLR in NLS1 may provide important 
clues to identify the parameter linking BLR and NLR 
excitation. 

Third, the spectra of NLS1 usually display strong high 
ionization forbidden emission lines (also known as
coronal lines) such as [Fe\,{\sc x}] $\lambda$6374, 
[Fe\,{\sc xi}] $\lambda$7892 
and [Fe\,{\sc xiv}] $\lambda$5303 (Penston et al. \cite{pe84}; 
De Robertis \& Osterbrock \cite{robo86}). Up to now,
it is not clear what the physical mechanisms involved in the
production of these features are, but three possible 
scenarios have been suggested: {\it (i)} collisional ionization  of 
gas with temperatures of T$_{\rm e} \sim$ 10$^{6}$ K (Oliva et al. 
\cite{oli94}); {\it (ii)} photoionization by the central 
non-thermal continuum (Ferguson et al. \cite{fkf97}); and 
{\it (iii)} a combination of shocks and photoionization 
(Viegas-Aldrovandi \& Contini \cite{vc89}; Contini \& Viegas 
\cite{cv01}). Although these lines are common in most AGN, 
NLS1 appears as suitable candidates to investigate their origin 
because they are not severely blended with
nearby broad features as in classical Seyfert 1 or diluted
by stellar absorption features as in Seyfert 2.  

The study of the NLR in NLS1 is not straightforward, 
however. Deblending the optical permitted lines in NLS1 is
difficult because no transition between the narrow and
broad components is observed. This shortcoming 
translates into large uncertainties in 
determining the fraction of H$\alpha$ and H$\beta$ that 
is emitted by the NLR, affecting in turn, the determination
of the reddening of the NLR gas. The extinction
correction should then relay on other indicators such as ratios of 
forbidden lines of the same ion, separated by a large wavelength
interval, yet not always observed with the same instrumental
setup. 

For the above reasons, we have started a program to 
study, in a consistent way, the NLR of a selected sample 
of NLS1 galaxies with the aim of investigating the physical 
conditions and structure of that region, with particular
emphasis in the origin of the coronal lines. To that purpose,
composite models accounting for both the effects of
photoionization by a central continuum source and shocks, 
created by collision of gas clouds 
with  the intercloud medium, will be employed.
The well-known NLS1 galaxy Arakelian\,564 ($z$ = 0.0247)
is a suitable candidate to begin with. The strong coronal 
line spectrum observed by Rodr\'{\i}guez-Ardila et al. 
(\cite{ro02}, hereafter RVPP02) in 
the near-infrared region (NIR) as well as the optical
measurements reported by Erkens et al. (\cite{erk97},
hereafter E97)
on this source form one of the most complete set of
lines available for modeling the NLR and studying the 
effects of shocks on line intensities in a NLS1. 
Following Moran (\cite{mor20}),  Ark\,564 exhibits
an unresolved radio  core and what appears to be a small scale
($\sim$ 1") jet.  This feature strengthens the hypothesis that shocks
are at work.
Moreover, it suggests that  extended synchrotron radiation should be
observed in the radio range.

Ark\,564 has been extensively studied from 
radio to X-rays, allowing a consistent modeling of 
both the line and continuum emission. Among other reasons 
that turns this target interesting is that it is 
the brightest known NLS1 in the 2-10 keV band (Collier et 
al. \cite{coll01}). The results of an intensive 
variability campaign on several wavelength bands show 
that the optical continuum is not significantly 
correlated with  the X-rays (Shemmer et al. \cite{shem01}).
The ionization state of the gas, as described by
Crenshaw et al. (\cite{cren02}), is relatively high.
They found that at least 85\% ~of the 
narrow emission-line flux comes from a region 
$\leq$ 95\,pc from  the nucleus and surrounded by a dust 
screen associated to a ``lukewarm'' absorber. 

In Sect.\,2 we describe the data set used to the study
the NLR. A description of the modeling output and
fitting to the line fluxes is given in Sect.\,3. As a
consistency check, the fit of the continuum 
emission SED is presented in Sect.\,4. 
Discussion of results follows in Sect.\,5 and Conclusions 
in Sect.\,6.

%__________________________________________________________________

\section{The data}

The NLR emission-line fluxes used in this work were primarily taken
from  E97 and RVPP02 for the 
optical (0.34-0.95\,$\mu$m) and  NIR (0.80-2.34\,$\mu$m)
lines, respectively.
Both sets of data were taken at 
similar resolution and S/N. Although the slit width 
employed in the optical observations was larger than
that of the NIR ones (1.5$''$ and 0.8$''$, 
respectively), the contamination introduced by the
host galaxy or extended emission to the optical fluxes is
negligible. In both cases, the authors report that most
of the integrated signal came from the unresolved
nuclear region and that the spectra were extracted from pixels
centered on the maximum of the continuum emission.
This is confirmed by comparing the flux of the 
[S\,{\sc iii}] 0.953\,$\mu$m line, common in both
spectra. E97 measured a value of 
3.06$\pm$0.13 $\times$ 10$^{-14}$ erg\,s$^{-1}$\,cm$^{-2}$, in
very good agreement with the value of 
3.11$\pm$0.10 $\times$ 10$^{-14}$ erg\,s$^{-1}$\,cm$^{-2}$
reported by RVPP02.

For reddening correction, we used the value of 
E(B-V)=0.09$\pm$0.03 reported by E97. 
Although it is a bit lower than the E(B-V)=0.14$\pm$0.04 
reported by Crenshaw et al. (\cite{cren02}), they agree
within errors. The former value was determined 
applying the method of Allen (\cite{all79}) to the [S\,{\sc ii}]
0.4068\,$\mu$m, 0.4076\,$\mu$m, [S\,{\sc ii}]
0.6717\,$\mu$m, 0.6731\,$\mu$m, [O\,{\sc ii}]
0.3727\,$\mu$m and [O\,{\sc ii}] 0.7330\,$\mu$m
line pairs, and so, most suitable to the NLR. 
Cols.\,2 and\,5 of Table~\ref{table1} list the extinction corrected 
line ratios relative to the total H$\beta$ flux.
In order to avoid the uncertainties introduced by 
the deblending of the observed permitted lines
into the component emitted by the NLR and BLR, 
we decided to normalize the narrow line fluxes to that
of [O\,{\sc iii}] 0.5007+0.4959\,$\mu$m. The resulting
values are reported in Cols.\,3 and\,6 of that same Table. 

\begin{table}
\caption[]{Emission line ratios$^{\mathrm{1}}$ for Ark\,564, corrected for E(B-V)=0.09}
\label{table1}
$$
\begin{array}{lccllcc}
\hline
\noalign{\smallskip}
Line     &   F_{\lambda}/H\beta\ &  F_{\lambda}/[\ion{O}{iii}] &  & Line     &   F_{\lambda}/H\beta\ &    %%@
F_{\lambda}/[\ion{O}{iii}] \\
\noalign{\smallskip}
\hline
\noalign{\smallskip}
 \lbrack\ion{Ne}{v}\rbrack\ \lambda3425 & 16.5\pm0.9 & 13.9\pm0.9 &  & \lbrack\ion{Fe}{x}\rbrack\ \lambda6374 &  %%@
7.10\pm0.42 & 5.97\pm0.41  \\
 \lbrack\ion{O}{ii}\rbrack\ \lambda3727 & 16.3\pm0.7& 13.7\pm0.7 &  & \lbrack\ion{N}{ii}\rbrack\ \lambda6548    & %%@
0.673\pm0.041 & 2.79\pm0.19 \\
 \lbrack\ion{Fe}{v}\rbrack\ \lambda3839 & 0.226\pm0.028 & 0.19\pm0.02 & & {\mathrm{H\alpha}}^{\mathrm{a}}       & %%@
355.0\pm14.7 & 298\pm16     \\
 \lbrack\ion{Ne}{iii}\rbrack\ \lambda3868& 9.75\pm0.41 & 8.2\pm0.4 & & \lbrack\ion{N}{ii}\rbrack\ \lambda6583 & %%@
9.97\pm0.44 & 8.30\pm0.46   \\
 \lbrack\ion{S}{ii}\rbrack\ \lambda4068 & 1.96\pm0.11 & 1.65\pm0.11 & & \ion{He}{i}~\lambda6678^{\mathrm{a}} & %%@
1.73\pm0.13 & 1.46\pm0.12  \\
 \lbrack\ion{S}{ii}\rbrack\ \lambda4076 & 0.482\pm0.038 & 0.40\pm0.03 & & \lbrack\ion{S}{ii}\rbrack\ \lambda6716 & %%@
3.88\pm0.32 & 3.26\pm0.29   \\
 \lbrack\ion{O}{iii}\rbrack\ \lambda4363 & 4.89\pm0.21 & 4.11\pm0.22 & & \lbrack\ion{S}{ii}\rbrack\ \lambda6730 & %%@
4.51\pm0.20 & 3.79\pm0.21    \\
 \ion{He}{ii}~\lambda4686^{\mathrm{a}} & 14.8\pm4.7 & 12.4\pm3.9  & & \lbrack\ion{Ar}{v}\rbrack\ \lambda7005    &  %%@
0.656\pm0.061 & 0.55\pm0.05    \\
 \lbrack\ion{Ar}{iv}\rbrack\ \lambda4711 & 1.01\pm0.09 & 0.85\pm0.08 & & \lbrack\ion{Ar}{iii}\rbrack\ \lambda7135 & %%@
1.69\pm0.09 & 1.42\pm0.09    \\
 {\mathrm{H\beta}}^{\mathrm{a}} & 100.0 & 84.1  &   & \lbrack\ion{Ar}{iii}\rbrack\ \lambda7751 & 0.538\pm0.100 & %%@
0.45\pm0.09   \\
 \lbrack\ion{Fe}{vii}\rbrack\ \lambda4893 & 0.573\pm0.075 & 0.48\pm0.06 &  & \lbrack\ion{Fe}{xi}\rbrack\ %%@
\lambda7891 & 8.31\pm0.39 & 7.0\pm0.40    \\
 \lbrack\ion{Fe}{vii}\rbrack\ \lambda4988 & 0.573\pm0.088 & 0.48\pm0.08 & & \lbrack\ion{S}{iii}\rbrack\ \lambda9069 %%@
& 5.76\pm0.26 & 4.80\pm0.27     \\
 \lbrack\ion{O}{iii}\rbrack\ \lambda5007+4959& 118.9\pm4.0 & 100.0 & & \lbrack\ion{S}{iii}\rbrack\ \lambda9532 & %%@
10.7\pm0.5 & 9.00\pm0.52 \\
 \lbrack\ion{Fe}{vii}\rbrack\ \lambda5145 & 0.283\pm0.023 & 0.23\pm0.02 & & \lbrack\ion{C}{i}\rbrack\ 0.985~\mu m & %%@
0.23\pm0.07 &  0.19\pm0.06    \\
 \lbrack\ion{Fe}{vi}\rbrack\ \lambda5159 & 1.01\pm0.09 & 0.85\pm0.08 & & \lbrack\ion{S}{viii}\rbrack\ 0.991~\mu m  %%@
& 1.75\pm0.13 & 1.47\pm0.12     \\
 \lbrack\ion{N}{i}\rbrack\ \lambda5199 & 2.04\pm0.12 & 1.72\pm0.12 & & \lbrack\ion{Fe}{xiii}\rbrack\ 1.074~\mu m  & %%@
2.88\pm0.55  & 2.42\pm0.47   \\
 \lbrack\ion{Fe}{vi}\rbrack\ \lambda5335 & 0.697\pm0.097 & 0.59\pm0.08 & & \lbrack\ion{S}{ix}\rbrack\ 1.252~\mu m & %%@
1.99\pm0.19 & 1.60\pm0.16 \\
 \lbrack\ion{Fe}{vii}\rbrack\ \lambda5721 & 2.31\pm0.15 & 1.94\pm0.14 & & \lbrack\ion{Fe}{ii}\rbrack\ 1.257~\mu m & %%@
0.97\pm0.16 &  0.82\pm0.14   \\
 \ion{He}{i}~\lambda5875^{\mathrm{a}} & 9.92\pm0.52 & 8.34\pm0.52 & & \lbrack\ion{Si}{x}\rbrack\ 1.430~\mu m    & %%@
4.55\pm0.32 & 3.83\pm0.30    \\
 \lbrack\ion{Fe}{vii}\rbrack\ \lambda6087 & 3.64\pm0.19 & 3.06\pm0.19 & & \lbrack\ion{Si}{vi}\rbrack\ 1.963~\mu m & %%@
1.89\pm0.35 & 1.59\pm0.30    \\
 \lbrack\ion{O}{i}\rbrack\ \lambda6300 & 5.29\pm0.23 & 4.45\pm0.24 & & \lbrack\ion{Ca}{viii}\rbrack\ 2.321~\mu m &  %%@
0.82\pm0.26 & 0.69\pm0.22     \\ 
\noalign{\smallskip}
\hline
\noalign{\smallskip}
\end{array} 
$$
\begin{list}{}{}
\item[$^{\mathrm{1}}$] Fluxes in the optical region, up to 8000 \AA, were taken from E97. NIR
lines, from 9000\AA\ and longward were taken from RVPP02.
\item[$^{\mathrm{a}}$] Permitted line with contribution from the BLR.
\end{list}
   \end{table}

\section{Modeling the spectra}

The main goal of this work is to model the NLR
spectrum of the NLS1 galaxy Ark\,564 using the fluxes of 
optical and NIR coronal lines (see Table~\ref{table1}) as 
major constraints. This ensures that
important information about high ionization material
(see Contini et al. \cite{cvp02a}) is
taken into account. The data from Table~\ref{table1} imply that
gas clouds with very different physical conditions must
coexist in the NLR of this object. This is reflected, 
e.g. in the large interval of critical 
densities (10$^{3}-$10$^{9}$ cm$^{-3}$) of the 
transitions and ionization
potentials (up to 370\,eV) of the ions involved in 
the production of the observed forbidden lines, or  
in the large range of line widths $-$a factor of 3 in
FWHM (see Table\,5 of RVPP02) measured in the line 
profiles. With this in mind, we have run a grid of 
models with parameters in the range suitable to the 
NLR of AGN and selected the models  that best 
matches  at least some of the observed line 
ratios. The grid was based on Contini \& Viegas 
(\cite{cv01}). The results are obtained by the SUMA code 
(Viegas \& Contini \cite{vc94}) which, besides
the line and continuum spectra emitted by 
the gas, also calculates re-radiation by
dust. 

Several multi-cloud models, obtained from a 
weighted average of single clouds models, were
calculated and compared with the observed data.
For each multi-cloud model, higher weights were 
assigned to single-cloud models matching the 
largest number of line ratios. Using this approach, 
we selected two final models that emerged as the best 
solutions. In the following sections they will be analyzed
in detail to determine which one best represents
the NLR state of Ark\,564.  

\subsection{The models}

The general model assumes  gaseous, dusty clouds, 
which move outwards from the active center (AC) and 
emit the line and continuum spectra.
A shock front forms on the outer edge of the clouds, 
while the inner edge is reached by the radiation 
from the AC.

The  input parameters are: the radiation flux intensity, 
\Fh~(in photons $\mathrm{cm^{-2}~s^{-1}~eV^{-1}}$ at 1 Ryd), 
the spectral index, $\alpha$ (= -1.5), the 
shock velocity, \Vs, the preshock density, \n0, the 
preshock magnetic field, \B0 (= $10^{-4}$ gauss), the 
dust-to-gas ratio by number, $d/g$, and the geometrical thickness, $D$, 
of the emitting clouds. Cosmic relative abundances (Allen 
\cite{all73}) are adopted. In radiation dominated (RD)   
clouds the effect of the flux from the AC prevails on the 
shock. In shock dominated (SD) clouds the flux is absent 
(\Fh = 0).

Notice that photoionization models with spectral index $\alpha$=-2
were adopted by  Rodr\'{\i}guez-Ardila et al. (\cite{ro00}) 
in order to explain the
low [\ion{O}{iii}]/H$\beta$ line ratio observed in the NLS1 galaxies.
The observations required both matter-bounded and ionization-bound
clouds. However no attempt was made to fit coronal lines or 
other emission lines
in wavelength ranges other than the optical one.  The presence of a
non-negligible velocity field, indicated by the FWHM of the emission-lines,
suggests that composite models, accounting for shock and photoionization,
and also providing regions of high and low density gas, are probably more
appropriate to model the emission-line spectra of NLS1 galaxies. 

Investigating the optical and UV properties of NLS1 galaxies, 
Constantin \& Shields (\cite{cons03}) suggest a spectral 
index of -1.34 in the UV--blue continuum. On the other hand,
{\it ROSAT} data show that the X-rays spectrum shows a power index steep and 
equal to -2.4 (Boller, Brandt \& Fink \cite{bbf96}), while ASCA data, 
corresponding to a different energy range and used to analyze the X-rays 
variability, indicate a power index of the order of -1.3.
Analyzing the optical spectrum of Ark\,564, Crenshaw et al. (\cite{cren02}) assume that the 
continuum of  Ark\,564 consists of power-laws with  index -1.0  
for energies less than 13.6 eV,
 -1.07 for 13.6 eV $\leq$ E $\leq$  0.5 keV, and
-1.6 for E $>$ 0.5 keV (consistent with Turner et al. \cite{tur01}).
Therefore, at least from the observational point of view,
there is no strong reason to adopt a power index steeper than
-1.6.

Here we adopt a UV -- X-rays power index 
$\alpha$ = -1.5 for all the models, recalling that the
shocked zone also contributes to the emission-line 
intensities, so that our results are less dependent
on the shape of the ionizing radiation.  
 As shown
in the next sections, our models with $\alpha$ = -1.5 provide a good fit to
the observed emission-line spectrum in a large range  of wavelengths.

In order to account for the constraints imposed by 
the observations (line widths, densities, etc.), the 
models selected from the grid show a 
large range of shock velocities (\Vs = 100 - 1500 \kms) 
and preshock densities (\n0 = 100 - 700 \cm3), as in
shown in Table~\ref{table2}. Notice that 
the shock velocities in the models lead to temperatures
in the post shock region between 1.5$\times$10$^5$ K (\Vs=100 \kms) 
and 3$\times$10$^7$ K (\Vs=1500 \kms), in agreement with the 
average temperature derived by E97 from the [\ion{Fe}{vii}] 
lines for the region producing the coronal lines. 
An overview of the values for 
the initial conditions adopted for each of the 14 individual
cloud models are listed in the first six entries of 
Table~\ref{table2}. Notice that the values $wr$ correspond to
the model relative weight. Thus, in order to obtain the 
fraction of the \Hb\ luminosity emitted by a given cloud, 
the relative weight should be multiplied by the 
corresponding \Hb\ value listed in Table~\ref{table2}
and normalized by the sum over all  clouds.

\begin{table}
\caption[]{Single-cloud models}
\label{table2}
\tiny{
\begin{tabular}{lllllllllllllll}\\ \hline\\
 & m1&m2&m3&m4&m5&m6&m7&m8&m9&m10&m11&m12&m13&m14\\ \hline\\
  \Vs$^1$ &100&100&100&100&150&150& 150&300&500&500&500&500&1000.&1500\\
  \n0$^2$ &100&100&100&100&30& 100&600&100&100&100&700&700&1000&300\\
  \Fh$^3$ &1.(9)&1.(10)&1.(11)&7.(11)&-&1.(12)&3.(12)&-&1.(10)&5.(12)&-&1.(11)&5.(12)&-\\
 D$^4$  &1.&3.&3.&3.&0.003&0.01&0.003&3.&3.&3.&3.&1.&0.1&3.85(-2)\\
 \Hb$^6$ %%@
&1.4(-2)&1.0(-1)&4.6(-1)&6.1(-1)1&1.6(-5)&3.3(-3)&3.7(-1)7&1.8(-4)&1.4(-3)&2.49(+1)&1.0(-2)&8.7(+3)&1.3(-2)&4.5(-2)%%@
\\
     & &&&&&&&&&&&&&\\
 Model M1 :&&&&&&&&&&&&&\\
 wr&4.7(-4)&1.9(-4)&0.0&2.4(-5)&9.5(-1)&4.7(-4)&9.5(-5)&0.0&3.8(-4)&1.9(-7)&1.3(-4)&2.(-8)&4.7(-2)&1.1(-4)\\
 d/g$^5$ &10.& 10.&1.&500.&1.&1.&1.&10.  &1000.&10.&1. &800.&-&-\\
 &&&&&&&&&&&&&&\\
\%(\lbrack\ion{Fe}{vii}\rbrack )&-&-&-&38.&21.&-&22.&-&0.7&9.&2.3&-&-&6.5\\
\%(\lbrack\ion{Fe}{x}\rbrack )&-&-&-&5.4&-&0.5&21.&-&-&1.&1.4&-&1.4&69.\\
\%(\lbrack\ion{Fe}{xi}\rbrack )&-&-&-&8.6&-&2.3&28.&-&-&1.5&0.9&-&8.2&50.\\
\%(\lbrack\ion{Fe}{xiii}\rbrack )&-&-&-&11.&-&18.&11.&-&0.5&1.4 &1.8&-&2.8&53.\\
\%(\lbrack\ion{C}{i}\rbrack )&-&-&-&-&28.&-&-&-&34.&-&38.&-&-&-\\
\%(\lbrack\ion{S}{iii}\rbrack )&4.&15.&-&0.9&31.&-&-&-&-&-&-&47.&-&-\\
\%(\lbrack\ion{S}{viii}\rbrack )&-&-&-&11.&-&0.6&58.&-&-&1.9&0.9&-&-&27.\\
\%(\lbrack\ion{S}{ix}\rbrack ) &-&-&-&6.7&-&1.7&23.&-&-&1.2&1.2&-&-&66.\\
\%(\lbrack\ion{Si}{vi}\rbrack )&-&-&-&40.&15.&-&29.&-&-&11.&0.9&2.1&-&1.2\\
\%(\lbrack\ion{Si}{x}\rbrack )&-&-&-&5.3&-&12.&20.&-&-&1.1&0.9&-&4.1&56.\\
\%(\lbrack\ion{O}{i}\rbrack )&23.&57.&-&-&12.&-&-&-&1.&-  &6.6&-&-&-\\
\%(\lbrack\ion{O}{ii}\rbrack )&13.&19.&-&-&65.&-&-&-&1.2&-&1.1&1.1&-&-\\
\%(\lbrack\ion{O}{iii}\rbrack )&1.2&11.&-&36.&34.&-&2.3&-&-&2.6&0.7&12.&-&-\\
\%(\lbrack\ion{Ne}{iii}\rbrack )&2.8&11.&-&21.&45.&-&0.8&-&-&16.&1.5&1.1&-&-\\
\%(\lbrack\ion{Ne}{v}\rbrack )&-&-&-&12.&22.&0.8&62.&-&-&23.&0.5&-&-&-\\
     & &&&&&&&&&&&&&\\
\ Model M2 :&&&&&&&&&&&&&&\\
\ wr&0.0&2.4(-4)&1.6(-5)&1.6(-4)&0.0&4.0(-2)&1.2(-3)&1.6(-1)&0.0&1.2(-5)&3.2(-3)&7.0(-7)&7.9(-1)&3.(-4)\\
\ &&&&&&&&&&&&&&\\
\%(\lbrack\ion{Fe}{vii}\rbrack )& -&-& -& 19.&-&-&20.&13.&-&42.&4.&-&-&1.4\\
\%(\lbrack\ion{Fe}{x}\rbrack )&-&-&-&5.3&-&6.2&39.&3.8&-&9.1&4.8&-&3.5&28.\\
\%(\lbrack\ion{Fe}{xi}\rbrack )&-&-&-&5.7&-&20.&35.&0.7&-&9.2&2.2&-&14.&14.\\
\%(\lbrack\ion{Fe}{xiii}\rbrack )&-&-&-&3.6&-&73.&7.&-&-&4.3&2.1&-&2.3&7.2\\
\%(\lbrack\ion{C}{i}\rbrack )&-&-&-&-&-&-&-&58.&-&-&42.&-&-&-\\
\%(\lbrack\ion{S}{iii}\rbrack )&-&1.1&0.5&-&-&-&-&2.&-&0.8&0.5&95.&-&-\\
\%(\lbrack\ion{S}{viii}\rbrack )&-&-&-&6.7&-&4.7&66.&3.1&-&11.&1.9&-&-&6.9\\
\%(\lbrack\ion{S}{ix}\rbrack ) &-&-&-&5.7&-&18.&37.&3.3&-&9.3&3.7&-&-&23.\\
\%(\lbrack\ion{Si}{vi}\rbrack )&-&-&-&18.&-&-&24.&4.8&-&47.&1.4&4.9&-&-\\
\%(\lbrack\ion{Si}{x}\rbrack )&-&-&-&2.2&-&63.&15.&0.7&-&4.2&1.3&-&4.2&9.6\\
\%(\lbrack\ion{O}{i}\rbrack )&-&26.&4.7&-&-&-&-&11.&-&-&57.&0.8&-&-\\
\%(\lbrack\ion{O}{ii}\rbrack )&-&3.8&0.6&-&-&-&-&85.&-&-&4.&6.2&-&-\\
\%(\lbrack\ion{O}{iii}\rbrack )&-&1.5&1.3&25.&-&-&3.&5.&-&17.&1.7&45.&-&-\\
\%(\lbrack\ion{Ne}{iii}\rbrack )&-&1.1&0.7&11.&-&-&0.8&6.1&-&75.&2.6&3.&-&-\\
\%(\lbrack\ion{Ne}{v}\rbrack )&-&-&-&7.&-&6.1&69.&3.8&-&13.&1.1&-&-&-\\
\hline\\
\end{tabular}}

Col.\,1: lines 1-4  show the input parameters, followed by the \Hb ~calculated
absolute flux for each model and then the output for the multi-cloud model M1
(relative weights wr, dust-to-gas ratios and percent contribution of each model to
each line); the same for the multi-cloud model M2 appears in the last lines of the table.
The $d/g$ results come from the SED which is modeled only for the selected model M1.

Cols.\,2-15 : the output of single model predictions.

$^1$ in \kms

$^2$ in \cm3

$^3$ in photons cm$^{-2}$ s$^{-1}$ eV$^{-1}$ at 1 Ryd

$^4$ in 10$^{19}$ cm

$^5$ in 10$^{-15}$

$^6$ in \erg

\end{table} 

The cloud geometrical thickness $D$ was chosen between 
3 and 10\,pc, but some models
represent filaments with $D$ $<$ 1\,pc. These filaments are created 
by fragmentation of matter in a turbulent regime
in the presence of shocks.
Such  clouds must be present in order to emit strong 
[Fe\,{\sc xi}] consistent with weak [C\,{\sc i}] lines.
The clouds emitting strong Fe coronal lines are 
matter-bounded, so they are mostly composed of highly 
ionized gas at high temperature. In a search for 
variations of [Fe\,{\sc vii}] and [Fe\,{\sc x}] in 
high ionization Seyfert galaxies, Veilleux 
(\cite{vei88}) claims that neither [Fe\,{\sc x}] nor 
[Fe\,{\sc vii}] in Ark\,564 show any 
significant variations between 1977 and 1988.
Considering that time variations are related to $D$/\Vs~ (the time
in which the shock crosses a filament), a  minimum time of
25 years would be needed  to detect any variability on
the line flux emitted by clouds of model m14 (see Table 2). 
This time would be much higher for other
clouds (see the values of $D$/\Vs), in perfect agreement with 
Veilleux's results. 

The  flux intensities range from 0 in SD models to 
\Fh = 5$\times$10$^{12}$  photons cm$^{-2}$ s$^{-1}$ eV$^{-1}$ at 1 
Ryd in RD models. The luminosity of Ark\,564 at 1 Ryd,  
L=7$\times$10$^{43}$ erg s$^{-1}$ (Romano et al. \cite{rom02}), 
constraints the models. Actually, the luminosity  at 1\,Ryd 
corresponding to the maximum \Fh= 5$\times$10$^{12}$ 
photons cm$^{-2}$ s$^{-1}$ eV$^{-1}$ is 4.5$\times$10$^{44} $ 
erg s$^{-1}$, assuming full coverage of the source.

The main criterion adopted for selecting the final
multi-cloud models is that the ratio of calculated 
results to  the data is within a factor of $\sim$ 2. 
However, given that the clouds are distributed over a
large volume of space, subjected to a variety of 
physical conditions, some line ratios may show slightly 
larger discrepancies. The single-cloud models which lead 
to the best two fitting multi-cloud models are presented in 
Table~\ref{table2} and labeled M1 and M2. The 
resulting line ratios normalized to 
[O\,{\sc iii}] $\lambda\lambda$5007+4959
are compared with the observed ratios in Table~\ref{table3}.

\begin{table}
\caption[]{Multi-cloud models results}
\label{table3}
\tiny{
$$
\begin{array}{llllllll}\\ \hline\\
 line &(I_{\lambda}/[\ion{O}{iii}])_{obs}&(I_{\lambda}/[\ion{O}{iii}])_{M1}& (I_{\lambda}/[\ion{O}{iii}])_{M2}  & %%@
RM1^1 & RM2^1 & RH\beta1^2 & RH\beta2^2 \\
\hline\\
\lbrack\ion{Ne}{v}\rbrack\, \lambda3425     & 140  & 250  & 290  & 1.77  & 2.10 & 2.32 & 1.33\\
\lbrack\ion{O}{ii}\rbrack\, \lambda3727     & 13.7 & 15   & 10   & 1.1   & 0.72 & 1.4  & 0.5\\
\lbrack\ion{Ne}{iii}\rbrack\, \lambda3868   & 8.2  & 7.0  & 9.6  & 0.84  & 1.17 & 1.1  & 0.74\\
\lbrack\ion{S}{ii}\rbrack\, \lambda4068 +   & 2.0  & 1.7  & 1.4  & 0.84  & 0.67 & 1.1  & 0.42\\
\lbrack\ion{O}{iii}\rbrack\, \lambda4363    & 4.0  & 3.9  & 1.7  & 0.94  & 0.41 & 1.24 & 0.23\\
\ion{He}{ii} \lambda4686                    & 12.0 & 22.0 & 40.0 & 1.80  & 3.19 & 2.38 & 2.0 \\
 \lbrack\ion{Ar}{iv}\rbrack\, \lambda4711   & 0.9  & 1.7  & 1.8  & 2.0   & 2.2  & 2.7  & 1.4\\
 \lbrack\ion{Fe}{vii}\rbrack\, \lambda4988  & 1.0  & 1.8  & 2.5  & 1.8   & 2.58 & 2.4  & 1.6\\
 \lbrack\ion{O}{iii}\rbrack\, \lambda5007   & 100  & 100  & 100  & 1.0   & 1.0  & 1.3  & 0.63 \\
 \lbrack\ion{N}{i}\rbrack\, \lambda5199     & 1.7  & 0.5  & 0.1  & 0.26  & 0.06 & 0.34 & 0.04\\
 \lbrack\ion{Fe}{vi}\rbrack\, \lambda5335   & 0.6  & 1.1  & 0.8  & 1.87  & 1.46 & 2.5  & 0.94\\
 \ion{He}{i} \lambda5875                    & 8.3  & 2.7  & 8.2  & 0.32  & 1.0  & 0.42 & 0.62\\
 \lbrack\ion{Fe}{vii}\rbrack\, \lambda6087  & 3.0  & 2.8  & 4.0  & 0.93  & 1.3  & 1.23 & 0.84\\
 \lbrack\ion{O}{i}\rbrack\, \lambda6300     & 4.4  & 4.5  & 1.3  & 1.01  & 0.30 & 1.33 & 0.20\\
 \lbrack\ion{Fe}{x}\rbrack\, \lambda6374    & 6.0  & 6.0  & 4.4  & 1.0   & 0.74 & 1.36 & 0.47\\
 \lbrack\ion{N}{ii}\rbrack\, \lambda6548+   & 9.0  & 21.0 & 7.2  & 2.3   & 0.8  & 3.0  & 0.50\\
 H\alpha                                    & 299  & 180  & 370  & 0.60  & 1.25 & 0.78 & 0.78\\
 \lbrack\ion{S}{ii}\rbrack\, \lambda6716    & 3.3  & 11   & 1.8  & 3.28  & 0.56 & 4.3  & 0.34\\
 \lbrack\ion{S}{ii}\rbrack\, \lambda6730    & 3.8  & 9.0  & 2.2  & 2.4   & 0.6  & 3.2  & 0.37\\
 \lbrack\ion{Ar}{iii}\rbrack\, \lambda7751  & 1.9  & 2.0  & 3.8  & 1.07  & 1.98 & 1.4  & 1.25\\
 \lbrack\ion{Fe}{xi}\rbrack\, \lambda7891   & 7.0  & 2.0  & 2.2  & 0.3   & 0.32 & 0.4  & 0.20\\
 \lbrack\ion{S}{iii}\rbrack\, \lambda9069+  & 14.0 & 10.0 & 19.0 & 0.71  & 1.36 & 0.94 & 0.86\\
 \lbrack\ion{C}{i}\rbrack\, 0.985\mu m      & 0.19 & 0.20 & 0.45 & 1.01  & 2.5  & 1.33 & 1.5\\
 \lbrack\ion{S}{viii}\rbrack\, 0.991\mu m   & 1.5  & 1.7  & 2.0  & 1.18  & 1.4  & 1.55 & 0.87\\
 \lbrack\ion{Fe}{xiii}\rbrack\, 1.074 \mu m & 2.4  & 2.3  & 5.0  & 0.96  & 2.0  & 1.3  & 1.3 \\
 \lbrack\ion{S}{ix}\rbrack\, 1.252 \mu m    & 1.7  & 0.9  & 0.7  & 0.52  & 0.43 & 0.68 & 0.27\\
 \lbrack\ion{Si}{x}\rbrack\, 1.430 \mu m    & 4.0  & 2.0  & 3.7  & 0.56  & 1.0  & 0.74 & 0.62\\
 \lbrack\ion{Si}{vi}\rbrack\, 1.963 \mu m   & 1.6  & 1.2  & 1.9  & 0.76  & 1.2  & 1.0  & 0.76\\
 \lbrack\ion{O}{iii}\rbrack/H\beta          & 1.19 & 1.56 & 0.75 &   -   &   -  &   -  &   -\\
 H\alpha/H\beta                             & 2.99 & 2.77 & 2.76 &   -   &   -  &   -  &   -\\
\hline\\
\end{array}
$$
}
$^1$ RM1 and RM2 stand for the ratio (I$_{\lambda}$/[\ion{O}{iii}])$_{calc}$/(I$_{\lambda}$/[\ion{O}{iii}])$_{obs}$ %%@
for M1 and M2, respectively.

$^2$ R\Hb1 and R\Hb2 stand for the ratio (I$_{\lambda}$/\Hb)$_{calc}$/(I$_{\lambda}$/\Hb)$_{obs}$ for M1 and M2, %%@
respectively.

\end{table}

The multi-cloud model M1 shows that [O\,{\sc iii}] is
overpredicted by a factor of $\sim$ 1.5
(compare the resulting ratio for \Ha) and
[Fe\,{\sc x}]/[Fe\,{\sc xi}] is higher than observed by a
factor of $>$2. Notice, however, that the observed lines
(E97, RVPP02) show  very complex line profiles,
which could be an indication that  more single-cloud
components should be included provided they do not
change significantly the overall spectrum emitted by the
cloud combination. However, in order
to include new components, additional constraints coming
from the line profiles should be accounted for, which
is out of the scope of this paper.
Moreover, model M1 gives [N\,{\sc ii}] and [S\,{\sc ii}]
values higher than observed. Since N and S are  not the
main gas coolants, this problem could be solved by changing
the adopted abundances of these two elements without
affecting the results for other lines.

On the other hand, model M2 shows a better agreement with 
the observed  [Fe\,{\sc x}]/[Fe\,{\sc xi}] ratio
($<$3), even if the [Fe\,{\sc xi}]/[O\,{\sc iii}] ratio 
is underpredicted, as well as  [O\,{\sc i}]/[O\,{\sc iii}], 
while the calculated He\,{\sc ii}/[O\,{\sc iii}] 
ratio  is high. In addition, [N\,{\sc ii}] and 
[S\,{\sc ii}] ratios are too low in this model, which would 
be improved using a higher abundance for N and S.

Both models M1 and M2 underpredict the 
[N\,{\sc i}]/[O\,{\sc iii}] ratio. Notice that the
[N\,{\sc i}] line is usually weak because the critical 
density for collisional deexcitation is very low 
($\sim$ 1000 \cm3) and the volume transition zone of the 
clouds is usually very low or inexistent.

Apart for the above discrepancies, it is clear from the 
data of Table~\ref{table3} that model M1 leads to a better 
agreement than M2 for a larger number of lines. But
before we can definitively choose  either of the two
models as the one that better represents  the NLR state
of Ark\,564, further constrains must be applied.  

\subsection{Relative contribution to the lines from different clouds -- 
weights and line widths}

The relative weights adopted to sum up the single-cloud 
models, $wr$, are  given in Table~\ref{table2}. The 
relative contribution (in per cent) of each cloud to the 
total predicted line flux  is also listed in Table~\ref{table2}
for the multi-cloud models M1 and M2.

The physical conditions of the gas in each cloud  are 
represented by the ensemble of the input parameters
of the corresponding model. In composite models,  the 
shock velocity and the flux from the AC are fundamental.
Although the shock velocity does not correspond exactly 
to the velocity of the gas, it can be used to discuss 
the FWHM of the lines presented by RVPP02 in their Table 5.
Clearly, all the line profiles show some substructure, 
confirming that many different conditions contribute to 
the different lines. Particularly, the FWHM are all 
$>$ 300 \kms. This is confirmed by our modeling, which 
shows (see Table~\ref{table2}) that all the coronal lines 
have strong contributions from clouds with \Vs\ $\geq$ 
300 \kms.

As an example, the [S\,{\sc iii}] line is mainly produced 
by clouds with \Vs=500 \kms\ (50\% in model M1 and 95\% 
for model M2), in good agreement with the value of 
FWHM = 440 \kms\ of RVPP02. The [S\,{\sc viii}] and 
[S\,{\sc ix}] lines, with more than 
50\% emitted by clouds with \Vs$<$ 300 \kms, have large wings 
with \Vs=1500 \kms. Regarding the [Fe\,{\sc vii}] line, it is
better reproduced by model M2 than by model M1.
In fact, model M2 shows that 60\% of that line  is emitted by 
clouds with \Vs=300-500 \kms, very close to the observed
FWHM of 420 \kms. 
[\ion{Fe}{x}] and [\ion{Fe}{xi}] are another set of
lines that deserves comments. Observationally, their 
line profiles have similar velocity distributions, with
FWHM of 570 \kms\ and 695 \kms, respectively. 
This result must be compatible with  the model outputs.
In fact, model  M1 predicts a large contribution ($\sim$60\%)
from high velocity clouds (\Vs $\geq$ 500 \kms) for both
lines. In contrast, according to model M2, [\ion{Fe}{x}] and 
[Fe\,{\sc xi}] are mostly emitted by low velocity clouds with 
\Vs=100 \kms\ and 150 \kms. The contribution of high velocity 
clouds, in this case, is not larger than 30\%. 

In addition to the line broadening produced by
the shock, bulk motions of the clouds around the gravitational 
potential of the central mass concentration should introduce
an additional broadening, not taken into account in our
modeling.  Although a detailed calculation of 
the resulting profile is out of the scope of this paper, it 
is important to note that our results, overall, are consistent with
the observed line widths. The reader should be aware that a 
complete modeling, able to simultaneously reproduce
line fluxes and widths as well as local variations in most
physical parameters, is not realistic because of the large number 
of free parameters, some of them unknown, that such
an approach would require. 

We conclude from the above discussion that model M1 is 
more representative of the actual conditions in the NLR of Ark\,564
than model M2. In addition to  better  fit the observed line
ratios, the predicted velocity distribution for the coronal gas
is more consistent with the observations.
The choice of model M1 is further reinforced by the discussion 
in the next section.

\subsection{Relative contribution of the BLR emission}

As it was pointed out in the Introduction Section, one of the
main problems associated to the study of the NLR in
NLS1 galaxies is the difficulty in obtaining a reliable
estimate of the permitted lines fluxes emitted by
that region. We recall here that the models calculated by 
SUMA refer to the NLR of Ark\,564 because the code is not 
adapted to calculations for emission from a very dense gas. 
However, comparing the observed and calculated line 
ratios relative to \Hb, as is shown in Cols.\,7 and\,8
of Table~\ref{table3} for models M1 and M2, respectively,
allows us to derived roughly the 
contribution of the broad \Hb. This is possible because
our permitted line flux measurements include the
contribution from both the NLR and BLR 
while the predicted line ratios, relative to \Hb, do not
include the broad line contribution. 
Thus, we expect that the calculated forbidden lines ratios,
relative to \Hb, be higher than the observed ones. 

This is in fact the situation found for the values 
predicted by model M1. They are $\sim$1.3 higher relative to
the ratios reported in Cols.\,2 and\,5 of Table~\ref{table1}, 
indicating that  the predicted \Hb$_{br}$ 
corresponds to about 70\% of the total observed \Hb.
On the other hand, for model M2 the values listed
in Col.\,8 of Table~\ref{table3} indicate the opposite, 
suggesting that this model is unrealistic.

This result supports Rodr\'{\i}guez-Ardila et al. (\cite{ro00}) 
conclusions about the fraction of the NLR
flux present in the observed H$\beta$ line of NLS1
galaxies. They found that, on average, 50\% of
the total integrated H$\beta$ flux is emitted by the NLR,
in contrast to the fraction of 10\% usually found in
broad-line Seyfert 1 galaxies. 

Notice that the ratio (I$_{\lambda}$/\Hb)$_{calc}$/(I$_{\lambda}$/\Hb)$_{obs}$
derived for \ion{He}{ii} $\lambda$4686/\Hb, \ion{He}{i} $\lambda$5876/\Hb, 
is not constant, indicating that the broad contribution  to 
\ion{He}{ii} $\lambda$4686, \ion{He}{i}, and \Hb\ is 
different. 

Finally, in Table~\ref{table5} we compare the results of model 
calculations with
the UV line ratios observed by Crenshaw et al. (\cite{cren02}), which 
include both the NLR and BLR contributions. Ratios
less (higher) than unity indicate that the contribution of the BLR to 
the line is high (low). 

\begin{table}
\caption[]{Comparison (I$_{\lambda}$/\Hb)$_{calc}$/(I$_{\lambda}$/\Hb)$_{obs}^1$
for the UV lines.}
\label{table5}
\begin{center}
\begin{tabular}{lc}\\ \hline\\
\     line    &  M1\\
\hline\\
 Ly$\alpha$ & 1.0\\  
 \ion{N}{v} 1214&0.33 \\
 \ion{O}{iv}\rbrack+\ion{Si}{iv} $\lambda$1400& 1.27\\
 \ion{N}{iv}\rbrack\,$\lambda$1486& 0.50\\
 \ion{C}{iv} $\lambda$1550& 1.23\\
 \ion{He}{ii} $\lambda$1640&  4.7\\
 \ion{O}{iii}\rbrack\,$\lambda$1663&0.66\\
 \ion{N}{iii}\rbrack\,$\lambda$1750&0.22 \\
 \ion{Si}{iii}\rbrack\,$\lambda$1890&1.60 \\
 \ion{C}{iii}\rbrack\,$\lambda$1909&1.19\\
 \lbrack\ion{Ne}{iv}\rbrack\,$\lambda$2423&1.36\\
 \ion{Mg}{ii} $\lambda$2800 & 0.20\\
\hline\\
\end{tabular}

$^1$ from Crenshaw et al. (\cite{cren02}, Table 2, reddening corrected)

\end{center}
\end{table}

\section{The SED of the continuum}

\begin{figure}
   \centering
   \includegraphics[width=10cm]{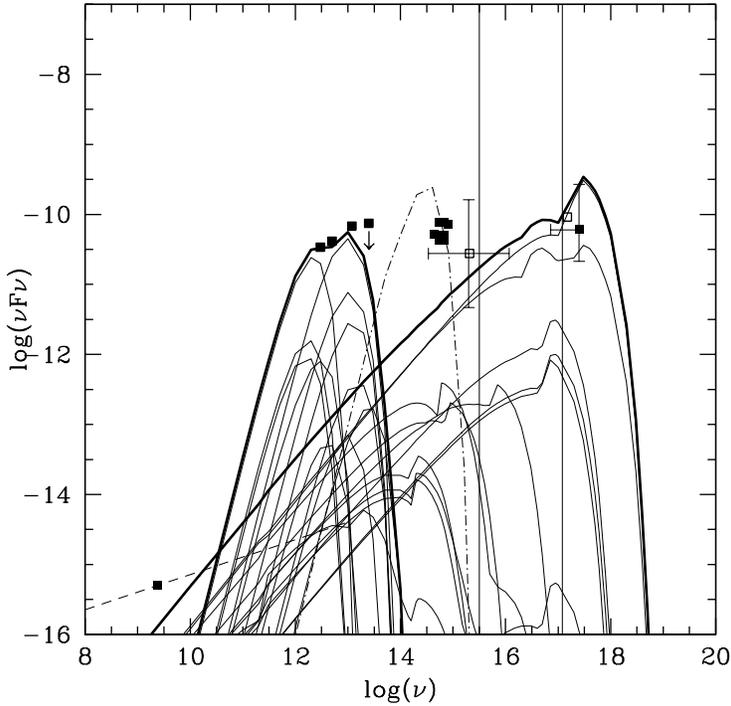}
      \caption{The observed continuum of Ark\,564 compared to the results of
the multi-cloud model M1. The observed data from the NED are shown as filled squares.
Open squares refer to the data of Walter \& Fink (1993) in the X-ray  and
of Rodr\'{\i}guez-Pascual et al. (\cite{rp97}) in the UV. 
The multi-cloud emission, corresponding to dust emission and bremsstrahlung
(thick solid lines) obtained from the weighted average of the clouds 
(thin solid lines) are shown, as well as the synchrotron emission from the 
shock front (dashed line). The old stellar population is represented by a 
blackbody (dot-dashed line). The frequency range where intrinsic absorption 
reduces the continuum flux is limited by the two thin vertical lines 
(see text).
              }
         \label{fig1}
   \end{figure}

Following the results from the previous sections,
model M1 is favored and its consistency with the observed continuum
in a large range of wavelengths is checked below.
In Fig.~\ref{fig1}
 the spectral energy distribution of the bremsstrahlung and of
re-radiation by dust from model M1 is compared
with the observed spectral energy
distribution of Ark\,564. 
The data were taken from the De Vaucouleur et al. (\cite{vanc1991}); 
Zwicky \& Kowal (\cite{zk1968}); De Vaucouleurs \& Longo (\cite{vl1988});
Moshir et al. (\cite{mos1990}) and Dressel \& Condon (\cite{dc1978}),
compiled in the NED database. The datum in the soft X-ray range 
come from the {\it ROSAT} All Sky Survey, reported by Walter \& Fink 
(\cite{wf1993}) and the observed flux at  
1450 \AA ~comes from {\it IUE} observations of Rodr\'{\i}guez-Pascual et al.
(\cite{rp97}).

%----------------------------------------------------------- 
%
%______________________________________________________________

Although the observed SED was constructed from data obtained at
very different epochs, it should be representative thanks to the
limited variability displayed by Ark\,564 
(Romano et al. \cite{ro02}). 

The most important physical
mechanisms that should contribute to the emission of the 
continuum in the different frequency ranges are
{\it (1)} bremsstrahlung from the photoionized gas 
($\nu$ $\sim$ 5$\times$10$^{13}$- 5$\times$10$^{15}$ Hz),
{\it (2)} bremsstrahlung from the  hot gas in the downstream region of
shocked clouds ($\nu$ $>$ 10$^{16}$ Hz), {\it (3)} emission 
from the old star population
($\nu$ $\sim$ 5$\times$10$^{13}$-10$^{15}$ Hz),
{\it (4)} dust re-radiation in the infrared range 
($\nu$ $\sim$ 10$^{12}$-10$^{14}$ Hz),
and {\it (5)} radio emission ($\nu$ $<$ 10$^{11}$ Hz).

Besides the line and continuum spectra, the SUMA code calculates
re-radiation by dust in a consistent way. Dust is  heated by
the nuclear radiation flux; moreover, gas and dust 
mutually heat each other by collisions
in shock turbulent regimes (Viegas \& Contini \cite{vc94}).
Therefore, the frequency corresponding to the dust 
re-radiation peak depends on the shock
velocity. At high \Vs\, the grains will reach high temperatures in the
post shock region and dust re-radiation will peak 
in the mid-IR, while at low velocities
the peak appears in the far-IR.
The peak intensity depends on the dust-to-gas ratio.
Sputtering of the grains, which depends on the shock velocity,
 is calculated throughout the clouds. Small grains are rapidly sputtered.
Therefore, we adopt silicate grains with an initial radius  of 0.2\,$\mu$m.

Line and continuum spectra must be modeled consistently.
The single-cloud models are calculated assuming a standard 
dust-to-grain, by number, $d/g$ = 10$^{-15}$,
providing a first fit to the emission lines. Once the best fit is reached
with a multi-cloud model, the continuum emission is checked and
the  $d/g$ values for the single-clouds may change. In this case,
an iterative process is carried out, in order to obtain a self-consistent
model accounting for the observed emission-line and continuum data.
Notice that an increase of $d/g$ in the postshock region, 
before complete sputtering
of the grains, acts as an increased density, speeding up the cooling
processes. However, an increase by a factor less than 50 changes the cloud
continuum emission without affecting the emission-line intensities.

The relative weights adopted to calculate the 
continuum SED are the same as those used to
model the line spectra. The dust-to-gas ratios 
(by number) that  fit the dust re-radiation
fluxes in the IR appear in  Table~\ref{table2}.
At high \Vs\, (models m13 and m14) the grains are completely
sputtered at the shock front and we cannot determine the $d/g$ ratio.
The dust-to-gas ratio, by number,  is $<$10$^{-13}$ in low velocity clouds
and as high as 8$\times$10$^{-13}$ in clouds with \Vs=500 \kms. Recall that
10$^{-14}$ by number corresponds to 4$\times$10$^{-4}$ by mass adopting silicate
grains with a density of 3 gr \cm3. 
It means that in Ark\,564 some high velocity clouds have  
dust-to-gas ratios higher than values found in the Galaxy.
The high $d/g$ found for some models (see Table~\ref{table2}) 
agrees to the relatively high
E(B-V) considered by Crenshaw et al. (\cite{cren02}).
They claim, in fact, that the continuum spectrum
of Ark\,564 is clearly much flatter than that of
Mrk 493 in the UV. This suggests that the spectrum
of Ark\,564 may experience a significant
amount of internal reddening, which is supported
by the \Ha/\Hb=3.4 measured by E97 and the small ratio
of UV to X-ray continuum flux observed in Ark\,564
compared to most Seyfert 1s (normal and NLS1s).

The two thin vertical lines in Fig.~\ref{fig1} show 
the range of frequencies corresponding to
strong absorption in the UV (see Contini et al. \cite{cvp02a}),
Within this interval, intrinsic absorption reduces 
the fluxes. Notice, for instance,  that the hydrogen 
column density is $>$ 10$^{22}$ cm$^{-2}$ for model m14. 
The data at $\nu \leq$ 10$^{15}$ Hz show that 
contribution from the old stellar
population is present in Ark\,564 (dash-dot line), as 
is found for most AGN. In the radio band, only one 
measurement in one frequency is available, so 
we cannot decide  whether there is also some contribution
from synchrotron emission created by Fermi mechanism 
at the shock front (dashed line, see 
Contini \& Viegas \cite{cv99}).

\section{Discussion}

The models which could better fit  the coronal 
lines of NLS1 galaxies, including Ark\,564, were already 
discussed in RVPP02 and they are
based on line ratios extracted from  a general grid of composite  models 
(Contini \& Viegas \cite{cv01}). The asymmetries
toward the blue observed in the line profiles of Ark\,564 (E97)
suggest that these lines originate in a
 gas outflowing from the nucleus, supporting the models that combine the
effect of shocks and photoionization. That first approach
was a start for a more detailed analysis of the NLR of these objects,
as presented in this paper. Here, emission-lines in a
 large range of wavelengths and ionization levels constrain
the modeling and reveal that many different conditions coexist
in the NLR.  Notice that multi-cloud models were used previously
in the detailed modeling of
NGC\,5252, Circinus, NGC\,4151, and NGC\,7130
(Contini et al. \cite{cv98a}, \cite{cv98b}, \cite{cvp02a}, 
\cite{crrr2}).

The multi-cloud models  in Ark\,564 show three main types of cloud velocities:
(a) Those between 100 and 150 \kms,
which are about the lowest observed in AGN and are typical of LINERS.
Very low velocities clouds,  with \Vs $\sim$ 60 \kms, have been
found to contribute  only to the
spectra of Circinus;
(b) Velocities of about 300-500 \kms, generally present in the NLR;
(c) High  velocity clouds ($\geq$ 1000 \kms), also
invoked in modeling  the Seyfert  galaxies referred above, 
have been used in order to explain high ionization lines ( e.g. [\ion{Fe}{x}])
as well as the soft X-ray emission. 
In the high \Vs ~clouds the {\it d/g} ratios are  higher
 than in  lower velocity  clouds by a factor of $\sim$ 100
both in Circinus and in NGC\,4151. 
 High velocity shocks lead to a strong heating of gas and dust. However,
the upper limit showed by Ark\,564 in
the  NIR continuum at log $\nu$ = 13.5  constrains
the $d/g$ in the high velocity clouds to lower values. The same
conclusion was obtained for  NGC\,5252.

Preshock densities \n0 $\leq$ 1000 \cm3 are found in Ark\,564, as
well as in the other previous modeled galaxies.
Only in the warm absorber  of NGC\,4051,
another NLS1, preshock densities up to 10$^6$ \cm3 were found in clouds  
characterized by very small $D$ ($\geq$ 2. 10$^{12}$ cm). 

Although the models  for Ark\,564 were chosen among those that
better fit the strongest lines, they should be able to
reproduce as much lines as possible. We briefly show the complex
modeling of such a rich spectrum, discussing for instance 
[\ion{Si}{x}]/[\ion{Si}{vi}]
relative to [\ion{Fe}{xi}]/[\ion{Fe}{x}] line ratios.
The observations show that [\ion{Si}{x}]/[\ion{Si}{vi}] $\geq$ 2 and 
[\ion{Fe}{xi}]/[\ion{Fe}{x}] $>$ 1.
Model m6, which is matter-bounded and RD, gives the right trend for the Fe
lines but strongly
underestimates the [\ion{Si}{vi}]/\Hb\ line ratio, which is otherwise high
in models composed of clouds with large $D$. In order to get a
consistent picture, the models must
be summed up
with a fine tuning of the relative weights  in order to reach an agreement
between observed and calculated ratios for all the lines.

Finally, if the NLR spectrum of Ark\,564 is well reproduced by composite
models accounting the effect of photoionization by a central source
and shocks, it is then natural to ask about the origin of such
shocks. Although SUMA does not discriminate about the
shock sources, the fact that Ark\,564 has an extended radio emission
makes that jet-induced shocks be a likely scenario. This hypothesis
is particularly suitable to explain the low-ionization lines. High velocity shocks, found to be adequate for coronal lines, may have
their origin from outflowing material that evaporates from
the torus. This last hypothesis, in fact, was already considered by
E97 and RVPP02 from their observations of coronal lines. 
In summary, the broad range of shock velocities found from our 
modeling is consistent with shocks generated, at least, from
two distinct mechanisms.

\section{Concluding remarks}

In the previous sections we have modeled, in detail, 
the observed continuum and line emission spectrum of the 
NLS1 galaxy Ark\,564. Particular emphasis is put on the 
constrains imposed by the coronal lines. This is the 
first time that such an approach is carried out for the NLR
of a NLS1 galaxy. The line spectrum ranges from UV to NIR 
wavelengths while the continuum extends from radio to
X-ray. Composite models accounting for the combined
effect of photoionization by a central source plus shocks
are used to that purpose. We found  a multi-cloud model that can suitably 
reproduce most of the observed line ratios within a factor of 2
and yields shock velocities consistent with the observed line widths.
The picture that emerges from our modeling is a NLR
composed of cloud showing a large range of
velocities (\Vs = 100 - 1500 \kms),
preshock densities (\n0 = 100 - 700 \cm3)
and illuminated by different flux intensities, from 0 
in SD models to \Fh = 5$\times$10$^{12}$ photons cm$^{-2}$ 
s$^{-1}$ eV$^{-1}$ at 1\,Ryd in RD models. The cloud 
geometrical thickness $D$ was found to be between
3 and 10\,pc, but some of them
represent filaments with $D <$ 1\,pc.
The single-cloud models are summed up adopting relative 
weights. The best fitting model (M1) is obtained by adopting 
the highest weights for SD clouds with \Vs=150 
and high velocity-density RD clouds  reached by a strong 
radiation flux. These high velocity  clouds, particularly   
SD clouds with \Vs = 1500 \kms\, account for the soft 
X-ray emission and for most of the flux from the  [Fe\,{\sc x}],  
[Fe\,{\sc xi}], and [Fe\,{\sc xiii}] lines
(70\%,  60\%, and 55\%, respectively). Clouds with
 \Vs=150 \kms\ and \n0=600 \cm3,
illuminated by the strong central continuum flux, also
contribute to the coronal lines. On the other hand, low ionization 
lines in the optical range are mostly emitted from SD clouds.
In the modeling process we have considered  forbidden line 
intensities relative to [O\,{\sc iii}], in  order to avoid  
the broad line contribution that is present in the permitted 
lines. 

The input continuum energy distribution in the 
UV--X-ray region that photoionizes
the NLR of Ark\,564 is characterized by a spectral index 
$\alpha$=-1.5.  This value is very close to the
$\alpha$ measured from UV and X-rays observations. 
That means that the NLR of Ark\,564 is directly illuminated 
by the radiation of the central source. In addition, shocks,
probably originated from the interaction between the
radio-jet and NLR gas, provides an additional input
mechanism for the production of coronal lines.

To complete the modeling, we have roughly evaluated 
the contribution of the BLR to the total \Hb ~line and to the  
permitted lines in the UV. A quantitative calculation 
is not possible because the SUMA code is not adapted 
to the calculation of lines from high density gas.
However, it is found that \Hb$_{broad}$/\Hb$_{narrow}$
should range between 1 and 2, in very good agreement
with the value found by Rodr\'{\i}guez-Ardila et al (\cite{ro00})
 for NLS1 galaxies
and determined from Gaussian fitting to the observed
permitted profiles.

Finally, we could consistently explain the data of the continuum 
by  the multi-cloud model M1. A gas-to-dust ratio between 
10$^{-15}$ and 10$^{-12}$ by number has been found for the NLR clouds,
showing  different values in  different clouds. 

\begin{acknowledgements}

We are very grateful to the referee, Dr. P. Ferruit,
for many interesting comments which improved the
presentation of the paper.
This paper is partially supported by the Brazilian agencies
FAPESP(00/06695-0) and CNPq (304077/77-1). This research has made use 
of the NASA/IPAC extragalactic database (NED), which is operated
by the Jet Propulsion Laboratory, California Institute of Technology,
under contract with the National Aeronautics and Space Administration.

\end{acknowledgements}

\end{document}